\documentclass[aip,jcp,reprint]{revtex4-1}

\usepackage{amsmath}
\usepackage{amssymb}
\usepackage{bm}
\usepackage{graphicx}
\usepackage{microtype}

\renewcommand{\vec}[1]{\bm{#1}}
\newcommand{\uvec}[1]{\bm{\hat{#1}}}
\newcommand{\avr}[1]{\left\langle#1\right\rangle}
\renewcommand{\d}{\operatorname{d}\!}

\DeclareMathOperator{\erfc}{erfc}

\begin{document}
\title{Trail-Mediated Self-Interaction}
\author{W. Till Kranz}
\email{kranz@thp.uni-koeln.de}
\affiliation{Institute for Theoretical Physics, Universit\"at zu
  K\"oln, Z\"ulpicher Stra\ss e 77, 50937 K\"{o}ln, Germany}
\affiliation{Rudolf Peierls Centre for Theoretical Physics, University
  of Oxford, Oxford OX1 3NP, United Kingdom}
\author{Ramin Golestanian}
\affiliation{Max Planck Institute for Dynamics and Self-Organization (MPIDS),
  Am Fa\ss berg 17, 37077 G\"ottingen, Germany}
\affiliation{Rudolf Peierls Centre for Theoretical Physics, University of Oxford, Oxford OX1 3NP, United Kingdom}

\date{\today}

\begin{abstract}
  A number of microorganisms leave persistent trails while moving along
  surfaces. For single-cell organisms, the trail-mediated self-interaction
  will influence the dynamics. It has been discussed recently [Kranz
  \textit{et al.} Phys. Rev. Lett. \textbf{117}, 038101 (2016)] that the
  self-interaction may localize the organism above a critical coupling
  $\chi_c$ to the trail. Here we will derive a generalized active particle
  model capturing the key features of the self-interaction and analyze its
  behavior for smaller couplings $\chi < \chi_c$. We find that fluctuations in
  propulsion speed shift the localization transition to stronger couplings.
\end{abstract}

\maketitle

\section{Introduction}
\label{sec:introduction}

Motility or active motion of an organism is most
useful if it can occur in response to external
stimuli.\cite{fenchel02,adler66,chet+mitchell76} It has been
recognized from the early days\cite{engelmann81} that directed motion
is already realized in both prokaryotic and eukaryotic single-celled
microorganisms.\cite{burrows12,rodiek+hauser15,maier+wong15} On the
microbial scale, chemical signals and the corresponding
response---chemotaxis---are the most widespread but by no means the
only stimuli that are used as information.

Surface dwelling microorganisms like the bacteria from the species
Pseudomonas, Neisseria and amoeboid slime molds are known to leave
trails\cite{merz+forest02,reid+latty12} of high molecular weight
biopolymers like
polysaccharides.\cite{burchard82,christensen+characklis90,zhao+tseng13}
It is believed that these trails are used as a means of cell-cell
communication in particular in the process of colony and spore
formation.\cite{bonner+savage47,harshey94,nakagaki01,higashi+lee07,kaiser07,bernitt+oettmeier10,reid+latty12,amselem+theves12,rodiek+hauser15}
The precise mechanisms, however, are still under active investigation.

Living organisms have intricate signal processing pathways, even on
the microbial scale.\cite{porter+wadhams11} Therefore the response to
a stimulus may be the result of a complicated control algorithm. On
the other hand there is evolutionary value in
robustness.\cite{devisser+hermisson03} Simplicity often lends itself
to robustness. In recent years there has been a growing recognition
that mechanistic models that forgo the intracellular chemical signal
processing may be able to explain surprisingly complicated
behavior.\cite{saha+golestanian14,palacci+sacanna15,gomez+blokhuis16,zoettl+stark16,illien+golestanian17}
Here we follow this \emph{active particle} approach and do not address
the question of how much internal signal processing is involved in
trail interaction.

Rapidly diffusing, small molecules such as cyclic adenosine
monophosphate (c\textsc{amp}) are also employed in cell-cell
communication in the form of
auto-chemotaxis.\cite{sengupta+vanteeffelen09,gelimson+golestanian15}
This form of inter-microbial communication is relatively well
understood both on the level of the intra-cellular signaling
path-ways and on the level of collective effects. Most artificial
active particles are auto-chemotactic in that they create and are
propelled by local chemical
gradients.\cite{illien+golestanian17,liebchen+loewen18} All these
systems are characterized by particles or organisms whose typical
speed $v_0$ or typical effective diffusivity $D$ is much smaller than
the diffusivity of the the signaling molecules $D_m \gg D,v_0R$ not 
least because their size $R$ is much bigger than molecular length
scales. In effect, an organism's self-generated concentration field is
to a lowest order approximation isotropic with respect to the organism 
itself even if the organism is in motion. Self-interaction of the organism 
with its own auto-chemotactic field due to direct coupling with 
the translational degrees of freedom has been studied in 
the past\cite{Tsori2004} and shown not to be strong enough 
to lead to trapping.\cite{Grima2005}

The trail material, on the other hand, consists of entangled
macromolecules\cite{sutherland01} with a very low diffusivity
$D_m\ll D,v_0R$. A moving microorganism will therefore encounter an
anisotropic distribution of its own trail. It obviously leaves a trail
\emph{behind} and not \emph{all around}. As a consequence, the only
observable dynamics is the effective dynamics that results from the
interplay of the organism's propulsion mechanism and the
trail-mediated self-interaction. We have recently shown that this
self-interaction may profoundly alter the
dynamics.\cite{kranz+gelimson16,gelimson+zhao16} Here we will discuss
this mechanism in more detail and on a more general basis.

We will start by specifying the model in Sec.~\ref{sec:bare-dynamics}
and derive the effective dynamics in
Sec.~\ref{sec:effective-dynamics}. Using this effective description we
will analyze the orientational dynamics in
Sec.~\ref{sec:angular-mean-square} and the translational diffusivity
in Sec.~\ref{sec:transl-mean-square}. We will briefly discuss the
influence of speed fluctuations in Sec.~\ref{sec:effect-speed-fluct}
before closing in Sec.~\ref{sec:conclusion}.

\section{Bare Dynamics}
\label{sec:bare-dynamics}

In the following we will be exclusively concerned with the dynamics of
a single microorganism on a pristine, essentially flat surface. The
state of a microorganism at time $t$ is fully described by its
position $\vec r(t)$, its orientation
$\uvec n(t) = (\cos\varphi(t), \sin\varphi(t))$ and the trail it has
left so far $\tilde\psi(\vec x, t)$. We assume active propulsion along
the current orientation $\uvec n(t)$ with a mean speed $v_0$ and
small fluctuations on top, characterized by a translational
diffusivity $D_v$,
\begin{equation}
  \label{eq:1}
  \d\vec r(t) = \uvec n(t)(v_0\d t + \sqrt{2D_v}\d W_t).
\end{equation}
Here $W_t$ denotes a Wiener process. Orientational persistence is
observed to be limited in microorganisms such that the rotational
diffusivity $D_r^0$ should be substantial. It has been found that a
torque is naturally generated by gradients of the trail field,
$\nabla\tilde\psi(\vec x, t)$, perpendicular to the microorganism's
instantaneous orientation. To be precise, the organism may not react to the
\emph{actual} trail field but only to the trail field it senses via
some transfer function $\Xi(\vec x, t)$ that may perform some spatial
averaging and potentially some time integration. The orientation
dynamics is therefore of the following form,
\begin{equation}
  \label{eq:2}
  \d\varphi(t) = \chi\nabla_{\perp}(t)(\Xi\ast\tilde\psi)(\vec r(t), t)\d t +
                 \sqrt{2D_r^0}\d Z_t,
\end{equation}
where $Z$ is a second, independent Wiener processes,\footnote{While
  there may be processes that generate fluctuations in translation and
  orientation at the same time\cite{gelimson+zhao16}, additional sources of noise and the
  coarse-graining of time to the diffusional time scale largely
  decorrelate the fluctuations.} $\chi$ characterizes the sensitivity
of the microorganism to the trail and
$\nabla_{\perp}(t) := \uvec e_{\varphi}(t)\cdot\nabla$ where
$\uvec e_{\varphi}(t)\perp\uvec n(t)$. The asterisk denotes a
convolution in space and time and $\Xi(\vec x, t)$ is normalized such
that $\int\d^2x\int\d t \;\Xi(\vec x, t) = 1$. Along its trajectory
$\mathfrak r(t) := \{\vec r(t')\}_{t'<t}$, the microorganism deposits
trail material with a constant rate $k$ and distributed according to a
compact profile $\tilde\Psi(x^2)$ normalized such that
$\int\tilde\Psi(x^2)\d^2x = 1$,
\begin{equation}
  \label{eq:3}
  \tilde\psi(\vec x, t|\mathfrak r(t))
  = k\int_{-\infty}^t\tilde\Psi\big([\vec x - \vec r(t')]^2\big)\d t'.
\end{equation}

Equations.~(\ref{eq:1}--\ref{eq:3}) constitute a set of stochastic
integro-differential equations for the time evolution of a single
crawling microorganism. We note that in order to understand the
dynamics of the microorganism, we do not need the full trail field
$\tilde\psi(\vec x, t)$ but only what the organism senses
$\nabla_{\perp}\psi(t) := \nabla_{\perp}(t)(\Xi\ast\tilde\psi)(\vec
r(t), t)$.
We imagine a primarily mechanic response to the trail such that the
transfer function is instantaneous on the time scales of the noise,
$\Xi(\vec x, t) \propto\delta(t)$. Then it is sufficient to consider
an effective profile $\Psi := \Xi\ast\tilde\Psi$, \textit{i.e.},
\begin{equation}
  \label{eq:4}
  \nabla_{\perp}\psi(t) = k\nabla_{\perp}(t)\int_{-\infty}^t\!\!\!\!\d t'
  \Psi\big([\vec r(t) - \vec r(t')]^2\big).
\end{equation}
We may then expand the gradient as
\begin{multline}
  \label{eq:5}
  \nabla_{\perp}(t)\Psi\big([\vec r(t) - \vec r(t')]^2\big) =\\
  = 2\uvec e_{\varphi}(t)\cdot[\vec r(t) - \vec r(t')]
  \Psi'\big([\vec r(t) - \vec r(t')]^2\big),
\end{multline}
where the prime on $\Psi$ denotes the derivative with respect to the
argument.

\section{Effective Dynamics}
\label{sec:effective-dynamics}

\begin{figure}[htb]
  \centering
  \includegraphics{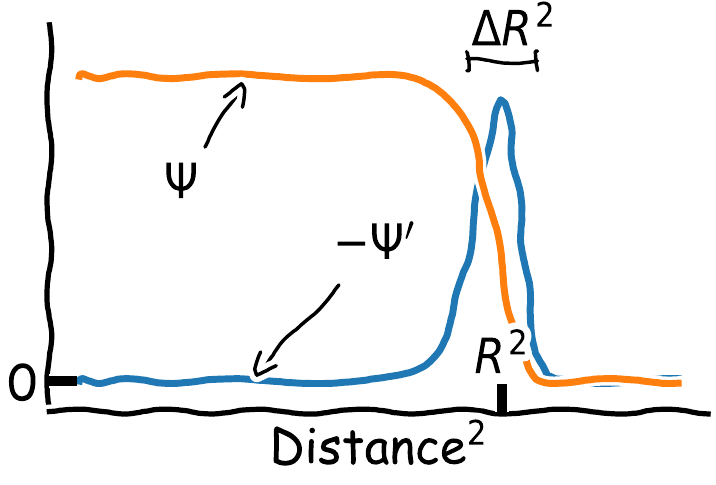}
  \caption{Sketch of the trail profile (as sensed by the organism) $\psi(r^2)$
    and its (negative) derivative $-\psi'(r^2)$ as a function of the squared
    distance from the trail's center. We assume well defined trail edges,
    $\Delta R/R\ll1$, throughout.}
  \label{fig:profile}
\end{figure}

To make progress we have to make a number of assumptions. We need the
trail profile to be sufficiently well defined (see
Fig.~\ref{fig:profile}) with a characteristic size $R$ and a trail
boundary of width $\Delta R\ll R$. Then we may approximate
$\Psi(x^2) \approx \Theta(R^2 - x^2)/\pi R^2$ where $\Theta(x)$ is the
Heaviside step-function. Likewise, we need the active propulsion speed
to be sufficiently well defined, \textit{i.e.}, $D_v\ll v_0R$. Then
the characteristic time to cross a trail $\tau$ is narrowly
distributed around $R/v_0$. We now assume \textit{a priori} that
trails are sufficiently straight that self crossings can be
neglected. We will find below that the self-interaction renormalizes
the rotational diffusivity $D_r^0$ to an effective value $D_r$,
\textit{i.e.}, we assume $D_r\avr\tau\ll1$ where $\avr{\tau}$ is the
mean crossing time. If the trajectories are straight enough so that the
organism only rarely crosses its own trail, we can safely ignore these
self-crossings.

In the following we will adopt units such that $R = v_0 = 1$ and
$k=\pi$. With the above assumptions, Eq.~(\ref{eq:1}--\ref{eq:3})
reduce to
\begin{align}
  \label{eq:8}
  \d\vec r(t) &= \uvec n(t)(\d t + \sqrt{2D_v}\d W_t)\\
  \label{eq:9}
  \d\varphi(t) &= \chi\nabla_{\perp}\psi(t)\d t + \sqrt{2D_r^0}\d Z_t\\
  \label{eq:10}
  \nabla_{\perp}\psi(t) &= \uvec e_{\varphi}(t)\cdot
  \int_{-\infty}^t\!\!\!\!\d t'\vec r_{tt'}\delta(1 - r_{tt'})
\end{align}
where $\vec r_{tt'} \equiv \vec r(t) - \vec r(t')$.

Due to the Dirac delta in Eq.~(\ref{eq:10}) and our assumption that
self-crossings are negligible, we only need to integrate over a time
interval of order unity. To this end, we iterate the equations of
motion (\ref{eq:8},\ref{eq:9}) to lowest order and find
\begin{equation}
  \label{eq:6}
  \begin{aligned}
    \vec r(t) - \vec r(t-t') &= \uvec n(t)t'
    + \sqrt{2D_v}\uvec n(t)W_{-t'}\\
    &+ \uvec e_{\varphi}(t)\int_t^{t-t'}\!\!\!\!\d u\int_t^u\times\\
    &\times\left[
      \chi\nabla_{\perp}\psi(w)\d w + \sqrt{2D_r^0}\d Z_w
    \right]\\
    &+ 2\sqrt{D_vD_r^0}\uvec e_{\varphi}(t)\int_t^{t-t'}\!\!\!\!Z_u\d W_u,
  \end{aligned}
\end{equation}
where we have used that
$\uvec e_{\varphi}(w) \approx \uvec e_{\varphi}(t)$ to lowest
order. In the following we are going to drop the last term in
Eq.~(\ref{eq:6}) because it contains the product of the small
parameters $D_v,D_r^0\ll1$.

The first term in Eq.~(\ref{eq:6}) turns the Dirac delta in
Eq.~(\ref{eq:10}) into a first passage time problem of a
one-dimensional Brownian motion with unit drift,
$t + \sqrt{2D_v}W_{t}$, on the positive half line with a reflecting
boundary at 0, \textit{i.e.},
\begin{equation}
  \label{eq:7}
  \int_{-\infty}^t\d t'\vec r_{tt'}\delta(1 - r_{tt'})
  = \vec r(t) - \vec r(t-\tau),
\end{equation}
where $\tau$ is a random variable, the first passage time. The
characterization of the first passage time distribution $p(\tau)$ is a
non-trivial task that we are not going to pursue here but note that
for $D_v\to0$, $p(\tau)\to\delta(\tau-1)$.

Making use of Eqs.~(\ref{eq:6}) and (\ref{eq:7}) in Eq.~(\ref{eq:10})
we find a closed equation for $\nabla_{\perp}\psi(t)$,
\begin{equation}
  \label{eq:12}
  \nabla_{\perp}\psi(t) = \int_t^{t-\tau}\!\!\!\!\!\!\d u\int_t^u\left[
    \chi\nabla_{\perp}\psi(w)\d w + \sqrt{2D_r^0}\d Z_w
  \right].
\end{equation}
Once we have a solution of Eq.~(\ref{eq:12}) we can use it in
Eqs.~(\ref{eq:8}, \ref{eq:9}) to analyze the effective dynamics of a
microorganism under trail-mediated self-interaction.

Let us start by considering a different representation of
Eq.~(\ref{eq:12}),
\begin{equation}
  \label{eq:13}
  \nabla_{\perp}\psi(t) = \int_0^{\tau}(\tau - w)\left[
    \chi\nabla_{\perp}\psi(t - w)\d w + \sqrt{2D_r^0}\d Z_{t-w}
  \right],
\end{equation}
to investigate the mean gradient
\begin{equation}
  \label{eq:14}
  \avr{\nabla_{\perp}\psi(t)}_Z = \chi\int_0^{\tau}\d w(\tau - w)
  \avr{\nabla_{\perp}\psi(t - w)}_Z.
\end{equation}
This is solved by the ansatz $\avr{\nabla_{\perp}\psi(t)}_Z \sim
e^{\alpha t}$ given the rate $\alpha$ solves the equation
$\lambda_{\tau}(\alpha) = 0$ where [cf. Eq.~(\ref{eq:21})]
\begin{equation}
  \label{eq:15}
  \lambda_{\tau}(\alpha) = 1 - \frac{\chi}{\alpha}\left[
    \tau + \frac1\alpha\big(e^{-\alpha\tau} - 1\big)
  \right].
\end{equation}
For vanishing speed fluctuations, $D_v\to0$, $\tau\equiv1$, and we
recover the behavior of Ref.~\onlinecite{kranz+gelimson16},
\textit{i.e.}, the average gradient relaxes to zero, $\alpha < 0$, for
weak coupling to the trail, $\chi < 2$, whereas it grows exponentially
above the critical value $\chi_c = 2$. For a discussion of the
localization transition that occurs for $\chi > \chi_c$ we refer to
Ref.~\onlinecite{kranz+gelimson16}.

For significant fluctuations, $D_v>0$, we need to analyze the ensemble
average,
$\lambda(\alpha) := \int_0^{\infty}\d\tau
p(\tau)\lambda_{\tau}(\alpha)$,
over the unknown first passage time distribution $p(\tau)$. For the
time being we assume that we may approximate
$\lambda(\alpha) \approx \lambda_{\avr\tau}(\alpha)$, \textit{i.e.},
by replacing the random variable $\tau$ by its mean which is known
exactly,\cite{mahnke+kaupusz09}
\begin{equation}
  \label{eq:11}
  \avr\tau = 1 + D_v\left(e^{-1/D_v} - 1\right) \simeq 1 - D_v.
\end{equation}
From $\lambda_{\avr\tau}(\alpha)$ we find a critical coupling strength
$\chi_c = 2/(1 - D_v)^2$ which is shifted to larger values for
increasing speed fluctuations.

For $\chi < \chi_c$, $\nabla_{\perp}\psi(t)$ represents a stochastic
process with zero mean and Eq.~(\ref{eq:12}) can be solved in the
Fourier
domain\footnote{$f(t) = \mathsf{FT}^{-1}[\tilde f](t) \equiv
  \int_{-\infty}^{\infty}\d\omega\tilde f(\omega)e^{i\omega t}$}
(cf.\ Sec.~\ref{sec:solving-eq-12})
\begin{equation}
  \label{eq:23}
  \widetilde{\nabla_{\perp}\psi}(\omega)
  = \frac{\sqrt{2D_r^0}}{\chi}\times
  [\lambda_{\tau}^{-1}(i\omega) - 1]i\omega\tilde Z_{\omega}.
\end{equation}
Note that by the Wiener representation theorem,
$i\omega\tilde Z_{\omega}\sim\mathcal N(0,1)$ are iid normal random
variables. In other words $\nabla_{\perp}\psi(t)$ is essentially
filtered white noise. In particular,
\begin{equation}
  \label{eq:24}
  \tilde\varphi(\omega) = \sqrt{2D_r^0}\lambda_{\tau}^{-1}(i\omega)
  \tilde Z_{\omega}
\end{equation}
is a stationary Gaussian process for $\chi < \chi_c$. This implies
that the joint probability,
\begin{equation}
  \label{eq:25}
  \begin{aligned}
    P(\varphi_1 - \varphi_2, t) &= P(\varphi_1, \varphi_2, t', t' + t)\\
    &= \frac{1}{\sqrt{2\pi\delta\varphi^2(t)}}
    e^{-(\varphi_1 - \varphi_2)^2/2\delta\varphi^2(t)},
  \end{aligned}
\end{equation}
is fully specified by the angular mean square displacement
\begin{equation}
  \label{eq:26}
  \delta\varphi^2(t) = \avr{[\varphi(t'+t) - \varphi(t')]^2}.
\end{equation}

\begin{figure*}[tb]
  \centering
  \includegraphics{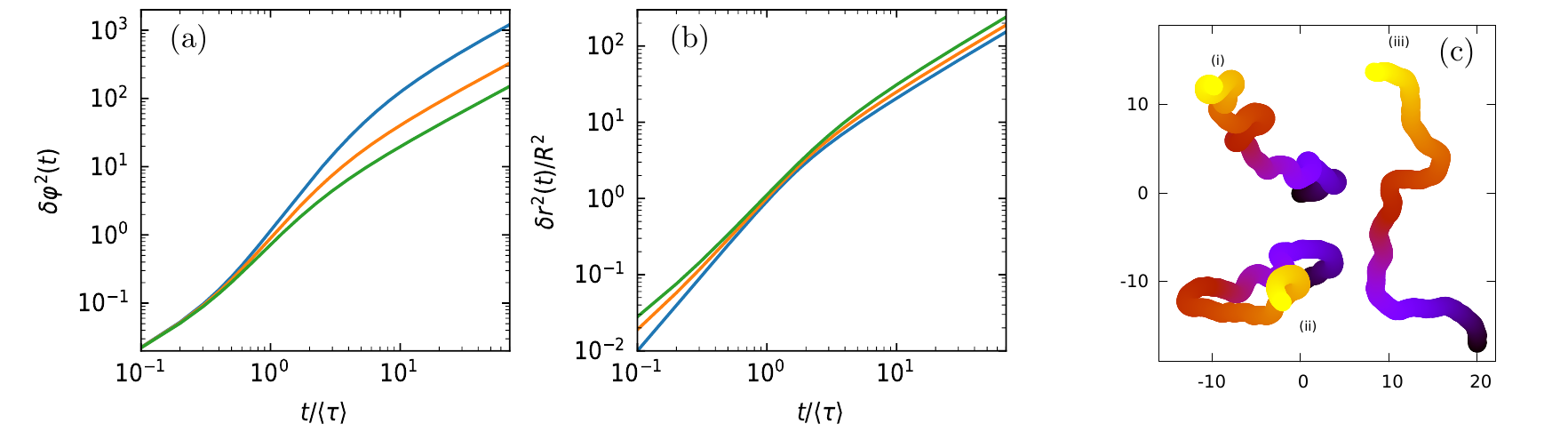}
  \caption{Influence of the speed fluctuations $D_v$ on the effective dynamics
    for strong coupling to the trail $\chi=1.8$. (a,b) Angular mean square
    displacement $\delta\varphi^2(t)$ and translational mean square
    displacement $\delta r^2(t)/R^2$ as a function of time $t$ normalized by
    the mean crossing time $\avr\tau$ for three different magnitudes of the
    speed fluctuations $D_v/v_0R = 0$ (blue), $0.05$ (yellow), $0.1$ (green)
    and a common value of the orientational diffusivity $D_r^0\avr\tau =
    0.1$. (c) Sample trajectories $\vec r(t)$ for the three different
    $D_v/v_0R = 0$ (i), $0.05$ (ii), and $0.1$ (iii) color coded as a function
    of time for a total duration $50\avr\tau$ and $D_r^0\avr\tau = 0.01$. See
    Sec.~\ref{sec:details-numerics} for details on the numerics.}
  \label{fig:msd}
\end{figure*}

\section{The Angular Mean Square Displacement}
\label{sec:angular-mean-square}

Using Eq.~(\ref{eq:24}) we find
\begin{equation}
  \label{eq:27}
  |s|^2\widehat{\delta\varphi^2}(s)
  = 2D_r^0\int_0^{\infty}\d\tau p(\tau)|\lambda_{\tau}(s)|^{-2},
\end{equation}
where $\hat f(s) = \mathsf{LT}[f](s) = \int_0^{\infty}\d tf(t)e^{-st}$
is the Laplace transform of $f(t)$,
$s = \sigma + i\omega\in\mathbb C$. To make progress and derive the
results discussed in Ref.~\onlinecite{kranz+gelimson16}, we will
again use the mean first passage time approximation,
\begin{equation}
  \label{eq:28}
  |s|^2\widehat{\delta\varphi^2}(s) = 2D_r^0|\lambda_{\avr\tau}(s)|^{-2}
\end{equation}
and will analyze $\lambda_{\avr\tau}^{-1}(s)$, cf.\ Eq.~(\ref{eq:15}).

In order to understand the angular mean square displacement in the
time domain, we need to know the analytical structure of
$\lambda^{-1}_{\avr\tau}(s)$.  Due to the oscillating factor
$e^{-i\omega}$ there are infinitely many poles in the complex plain and
an analytical inverse Laplace transform is not tractable. We may,
however, consider the asymptotic limits $t\to\infty$ and $t\to0$. Note
that, apart from the trivial scale factor $D_r^0$, the only control
parameters that affect $\delta\varphi^2(t)$ is the coupling strength
$\chi$ and the speed fluctuations $D_v$.

\subsection{Short-Time Asymptotics}
\label{sec:short-time-asympt}

Expanding $\lambda_{\avr\tau}(\sigma)$ in powers of $\sigma^{-1}$ we find
\begin{multline}
  \label{eq:29}
  \lambda_{\avr\tau}(\sigma) = 1 - \chi(1 - D_v)\sigma^{-1} + \chi\sigma^{-2} \\
  + O\big(\sigma^{-2}e^{-1/\sigma^{-1}}\big) .
\end{multline}
The first three terms are dominant as long as $e^{-1/\sigma^{-1}}\ll1$,
\textit{i.e.}, as long as $\sigma^{-1}\ll1$, or upon reinstating units, for
times $t\ll\avr{\tau}$. Then we have
\begin{multline}
  \label{eq:30}
  |\lambda_{\avr\tau}(\sigma\to\infty)|^{-2}
  \simeq 1 + 2(1 - D_v)\chi\sigma^{-1}\\
  - 2\chi(1 + \chi/\chi_c)\sigma^{-2},
\end{multline}
and in the time domain
\begin{equation}
  \label{eq:31}
  \delta\varphi^2(t\to0)/2D_r^0 \simeq t + \chi(1 - D_v)t^2
  - \frac13\chi(1 + \chi/\chi_c)t^3.
\end{equation}
In other words, the angular mean square displacement starts with the
bare diffusivity $D_r^0$ before the self-interaction becomes visible
on times of the order $1/\chi(1 - D_v)$.

\subsection{Long-Time Asymptotics}
\label{sec:long-time-asympt}

Expanding $\lambda_{\avr\tau}(\sigma)$ in powers of $\sigma$ we find
\begin{equation}
  \label{eq:34}
  \lambda_{\avr\tau}(\sigma) = 1 - \chi/\chi_c
  + \frac{\chi}{3\chi_c}(1 - D_v)\sigma + O(\sigma^2).
\end{equation}
This shows that $\delta\varphi^2(t\to\infty) = 2D_rt$ will
asymptotically always be diffusive with a renormalized
diffusivity\footnote{This corrects a misprint in Eq.~(4) of
  Ref.~\onlinecite{kranz+gelimson16}}
\begin{equation}
  \label{eq:35}
  D_r/D_r^0 = |\lambda_{\avr\tau}(0)|^{-2}
  = 1 + \frac{\chi}{\chi_c}\times\frac{2 - \chi/\chi_c}{(1 - \chi/\chi_c)^2}
\end{equation}
which diverges for $\chi\to\chi_c$.

The validity of the long time asymptotics is bounded by the radius of
convergence of the Taylor expansion, Eq.~(\ref{eq:34}). The latter is
determined by the location of the pole of $\lambda_{\avr\tau}^{-1}(s)$
closest to the origin of the complex plane. In
Sec.~\ref{sec:poles-near-origin} we show that Eq.~(\ref{eq:34}) holds
for $|s| < 3(\chi_c/\chi - 1)/\avr\tau$, \textit{i.e.}, for times
\begin{equation}
  \label{eq:48}
  t/\avr\tau \gg t^* := \frac{\chi/\chi_c}{1 - \chi/\chi_c}.
\end{equation}
The onset of the asymptotic regime, $t^*$, diverges with the rotational
diffusivity as $\chi\to\chi_c$.

For $\chi\to\chi_c$ we write $\chi = \chi_c(1 - \delta\chi)$. Assuming the
smallest pole $\sigma = O(\delta\chi)$ to be confirmed below we
expand to lowest order
\begin{equation}
  \label{eq:37}
  \lambda_{\avr\tau}(\sigma) = \delta\chi + \frac13(1 - D_v)\sigma
  + O(\delta\chi^2, \sigma^2, \sigma\delta\chi).
\end{equation}
Close to the critical coupling strength we therefore find
\begin{equation}
  \label{eq:38}
  |\lambda_{\avr\tau}(\sigma)|^{-2} \simeq \frac{9}{(1 - D_v)^2\sigma^2}
  \times\frac{1}{[1 + 3\delta\chi/(1-D_v)\sigma]^2},
\end{equation}
\textit{i.e.}, a superballistic behavior
\begin{equation}
  \label{eq:42}
  \delta\varphi^2(t) = \frac32\chi_cD_r^0t^3
\end{equation}
in a diverging time window $\avr\tau \ll t \ll 1/3\delta\chi$.

\section{The Translational Mean Square Displacement}
\label{sec:transl-mean-square}

The velocity autocorrelation function
$C(t) := \avr{\dot{\vec r}(t+t')\cdot\dot{\vec r}(t')} = 2D_v\delta(t)
+ C_{nn}(t)$
where $C_{nn}(t) := \avr{\uvec n(t+t')\cdot\uvec n(t')}$ can be
determined explicitly with the help of Eq.~(\ref{eq:25}),\cite{doi+edwards88}
\begin{equation}
  \label{eq:32}
  C_{nn}(t) = \int_{-\infty}^{\infty}\d\varphi P(\varphi, t)\cos\varphi
  = e^{-\delta\varphi^2(t)/2}.
\end{equation}
The asymptotic translational diffusivity is then given by a Green-Kubo
integral
\begin{equation}
  \label{eq:39}
  D/D^0 = D_r^0\int_0^{\infty}\d t\left[
    2D_v\delta(t) + e^{-\delta\varphi^2(t)/2}
  \right],
\end{equation}
where $D^0 = 1/D_r^0$ is the diffusivity for
$\chi = D_v = 0$.\cite{doi+edwards88}

The (translational) mean square displacement
$\delta r^2(t) := \avr{[\vec r(t+t') - \vec r(t')]^2}$,
\begin{align}
  \label{eq:33}
  \delta r^2(t) &= 2\int_0^t\d u\int_0^u\d wC(w)\\
  &= 2D_vt + 2\int_0^t\d w(t - w)e^{-\delta\varphi^2(w)/2}
\end{align}
cannot be given in closed form. However, the form of Eq.~(\ref{eq:33})
indicates, that it will be dominated by the small angle behavior,
$\delta\varphi^2(t) < 1$, of the angular mean square displacement.

In the following we will derive analytic expressions for the
diffusivity $D$ in certain limiting cases.

\subsection{Short Persistence Regime}
\label{sec:short-pers-regime}

For parameters such that $\delta\varphi^2(\avr\tau)\gg1$, we may use the
short time expansion, Eq.~(\ref{eq:31}). With this, the condition
reads $2D_r^0\gg1/[1 + \chi(1 - D_v)]$ which shows that this regime
applies for small coupling strength $\chi$ and large intrinsic
rotational diffusivity $D_r^0$. Given this is fulfilled, we have
\begin{align}
  D/D_0 &= D_v/D^0 + D_r^0\int_0^{\infty}\d t
          e^{-D_r^0[t + \chi(1 - D_v)t^2]}\\
  &= \sqrt{\pi/\kappa}\erfc\big(1/\sqrt{\kappa}\big)e^{1/\kappa} +
    D_v/D^0\\
  \label{eq:40}
  &= 1 + D_v/D_0 - \kappa/2 + 3\kappa^2/4 + O(\kappa^3),
\end{align}
where $\kappa := 4\chi( 1 - D_v)/D_r^0\ll1$ is a kind of Peclet number
relating the ``convective'' rate $\chi$ to the diffusive rate $D_r^0$.

\subsection{Long Persistence Regime}
\label{sec:long-pers-regime}

The opposite limit is given by an angular mean square displacement
which reaches the value one well into the asymptotic regime,
$\delta\varphi^2(t^*) \ll 1$. A condition which may be estimated as
$D_r/(1 - \chi/\chi_c) \ll 1$. Then we may approximate
$\delta\varphi(t) = 2D_rt$ for all relevant times and directly find
\begin{equation}
  \label{eq:41}
  D/D^0 = D_r^0/D_r,\text{ or, equivalently, }D = 1/D_r.
\end{equation}
In other words for very persistent trails, \textit{i.e.}, small
intrinsic directional noise $D_r^0$ and/or small coupling strength
$\chi\ll\chi_c$, the asymptotic translational diffusivity is inversely
proportional to the asymptotic rotational diffusivity. Higher order
terms are given in Ref.~\onlinecite{kranz+gelimson16}.

\subsection{Critical Regime}
\label{sec:critical-regime}

Close to the critical coupling strength, $\chi\to\chi_c$, we use
Eq.~(\ref{eq:42}), to make the ansatz
\begin{equation}
  \label{eq:43}
  \delta\varphi^2(t)/2D_r^0 = t\Theta(\avr\tau - t)
  + 3\chi_ct^3\Theta(t - \avr\tau)/4,
\end{equation}
patching together the short time, bare diffusion $2D_r^0t$, and the
intermediate time, superballistic behavior, $\propto t^3$. The
asymptotic diffusion is irrelevant here because
$\delta\varphi^2(t)\gg1$ before it sets in. Using this ansatz in
Eq.~(\ref{eq:39}), we find
\begin{multline}
  \label{eq:44}
  D/D^0 = D_v/D^0 + D_r^0\int_0^{\avr\tau}\d te^{-D_r^0t}\\
  + D_r^0\int_{\avr\tau}^{\infty}\d te^{-3D_r^0\chi_ct^3/4}.
\end{multline}
The second integral can be expressed in terms of the generalized
exponential integral $E_n(x)$,
\begin{multline}
  \label{eq:16}
   D/D^0 = 1 + D_v/D^0 - e^{-D_r^0\avr\tau}\\
   + \frac13D_r^0\avr\tau E_{2/3}(3D_r^0\avr\tau/2).
\end{multline}
Consistent with our assumption $D_r^0\avr\tau\ll1$ we need to expand
this to yield
\begin{equation}
  \label{eq:45}
  D/D^0 \simeq \Gamma(4/3)(D_r^0\avr\tau)^{2/3}
  - (\sqrt[3]{3/2}\avr\tau - 1)D_r^0,
\end{equation}
where $\Gamma(x)$ is the Euler Gamma-function. Note that for a
perfectly persistent organism, $D_r^0\to0$, the translational
diffusivity will obviously diverge $D \sim (D_r^0\avr\tau)^{-1/3}$.

\section{The Effect of Speed Fluctuations}
\label{sec:effect-speed-fluct}

Nonzero speed fluctuations $D_v > 0$ have multiple effects as can bee
seen by the examples in Fig.~\ref{fig:msd}. For the angular mean
square displacement, the influence lies mostly in the distance to the
critical point $\chi_c$. At vanishing fluctuations, $D_v = 0$, the
chosen coupling strength $\chi= 1.8$ is close to the critical value
$\chi_c = 2$ and the trajectories already begin to violate the
assumptions of the derivation by turning quickly. For increasing
fluctuations $D_v$, the critical value $\chi_c = 2/(1 - D_v)^2$
shifts to higher values. In effect, both the intermediate regime of
the angular mean square displacement as well as the asymptotic
diffusivity $D_r$ decrease and the trajectories become straighter.

For the translational mean square displacement $\delta r^2(t)$, the effects
are less drastic but can be seen in both the short and the long time
limit. For short times, the ballistic regime, $\delta r^2(t) \propto t^2$ is
replaced by diffusive behavior, $\delta r^2(t) = D_vt$. At intermediate times,
the directed motion prevails if the short time diffusivity is small enough as
it has been discussed by \citet{peruani+morelli2007}. For long times the
straighter trails enhance translational diffusivity for increasing
fluctuations $D_v$ but rather mildly so because the differences in
$\delta\varphi^2(t)$ mostly occur at large displacements
$\delta\varphi^2(t)\gg1$.

\section{Conclusion}
\label{sec:conclusion}

We have started by motivating a model of a self propelled particle
(the microorganism) on a two-dimensional plane that interacts with its
own trail, cf.\ Eqs~(\ref{eq:1}--\ref{eq:3}). A simplified version of
this model has been introduced by us before.\cite{kranz+gelimson16}
Here we argued that in reality the microorganism will not interact with
the trail, $\tilde\psi$, itself but with its observation of the trail,
$\psi$. Given that $\psi$ has well defined edges, $\Delta R/R\ll1$,
that the trails are reasonably straight, $D_r\avr\tau\ll1$ and the
propulsion speed $v_0$ is well defined, $D_v/Rv_0\ll1$, we showed how
to decouple the equation for the trail's gradient, Eq.~(\ref{eq:12}),
from the equation of motion of the particle,
Eqs.~(\ref{eq:8},\ref{eq:9}).

Analyzing the effective trail gradient, Eq.~(\ref{eq:12}), we showed
that it fails to regress to a zero mean beyond a critical coupling
strength
\begin{equation}
  \label{eq:36}
  \frac{k\chi_c}{\pi v_0^2R} = \frac{2}{(1 - D_v/v_0R)^2} \geq 2.
\end{equation}
However for $\chi < \chi_c$, both the effective trail gradient,
Eq.~(\ref{eq:23}), as well as the orientation, Eq.~(\ref{eq:24}), turn
out to be filtered white noise. The filter function
$\lambda_{\tau}^{-1}(i\omega)$, Eq.~(\ref{eq:15}), therefore, is
crucial for the dynamics. In essence we derived generalized
expressions for the effective angular [Eq.~(\ref{eq:35})] and
translational diffusivity [Eqs.~(\ref{eq:40}, \ref{eq:41},
\ref{eq:45})], special cases of which have been presented in
Ref.~\onlinecite{kranz+gelimson16}.

The effect of the self-interaction becomes apparent on a timescale $t/\avr\tau
\sim 1/\chi$ which indicates the start of an intermediate regime displaying
angular superdiffusion that extends to times $t/\avr\tau \sim \chi/\chi_c(1 -
\chi/\chi_c)$ beyond which the dynamics is effectively diffusive again. The
onset of the asymptotic regime diverges for $\chi\to\chi_c$. The translational
dynamics is unaffected by the asymptotic behavior of the angular motion due to
the exponential suppression in Eq.~(\ref{eq:32}). Consequently, the
translational diffusivity remains finite at the critical coupling
$\chi\to\chi_c$, Eq.~(\ref{eq:45}). A detailed analysis of the localized 
phase $\chi > \chi_c$ requires a new approach that includes 
the effect of frequent self-crossings neglected here and is left 
to future work.

We note that trail-mediated self-interactions have been recently studied 
in the context of aggregation and collective motion of myxobacteria, 
where experiments revealed that the collective motion of the wild type 
of Myxococcus Xanthus is organized in a network structure made out 
of trails, in contrast with signaling-deficient mutants of 
the species.\cite{Starruss2012,Thutupalli2015} These results have 
been rationalized theoretically using a phenomenological agent-based 
model with implemented trail-following rules.\cite{Balagam2015} 
It will be interesting to extend our study to mechanistically study 
the case of myxobacteria by incorporating both positional and 
orientational coupling to the trail and making predictions about 
the collective behaviour as as been performed for the case 
of Pseudomonas aeruginosa.\cite{kranz+gelimson16,gelimson+zhao16}
We also note that the self-trapping reported here is somehow reminiscent 
of the milling patterns that appear in a similar cognitive flocking model 
of animals.\cite{Peruani2016}

\begin{acknowledgments}
  We acknowledge helpful discussions with Anatolij Gelimson. This
  work was supported by the Human Frontier Science Program
  RGP0061/2013. W.\ T.\ K.\ thanks the \textsc{dfg} for partial
  funding through KR\ 4867/2-1.
\end{acknowledgments}

\appendix

\section{Solving Eq.~(\ref{eq:12})}
\label{sec:solving-eq-12}

To this end we start from yet another representation of
Eq.~(\ref{eq:12}),
\begin{multline}
  \label{eq:17}
  \nabla_{\perp}\psi(t) = \sqrt{2D_r^0}\int_0^{\tau}\d u
  (Z_t - Z_{t-u})\\
  + \chi\int_0^{\tau}\d u\int_{t-u}^t\d w\nabla_{\perp}\psi(w).
\end{multline}
Employing the Fourier representation, we find for the first term
\begin{multline}
  \label{eq:18}
  \int_0^{\tau}\d u(Z_t - Z_{t-u}) =\\
  = \int_{-\infty}^{\infty}\!\!\d\omega\tilde Z_{\omega}e^{i\omega t}
  \int_0^{\tau}\d u(1 - e^{-i\omega u})
\end{multline}
and for the second term
\begin{multline}
  \label{eq:19}
  \int_0^{\tau}\d u\int_{t-u}^t\d w\nabla_{\perp}\psi(w) =\\
  = \int_{-\infty}^{\infty}\!\!\d\omega\int_0^{\tau}\d u
  \int_{t-u}^t\d w\widetilde{\nabla_{\perp}\psi}(\omega)e^{i\omega w}.
\end{multline}
Upon performing the $w$-integral this yields
\begin{multline}
  \label{eq:20}
  \int_0^{\tau}\d u\int_{t-u}^t\d w\nabla_{\perp}\psi(w) =\\
  \int_{-\infty}^{\infty}\!\!\d\omega
  \frac{\widetilde{\nabla_{\perp}\psi}(\omega)}{i\omega}e^{i\omega t}
  \int_0^{\tau}\d u(1 - e^{-i\omega u}).
\end{multline}
Together with
\begin{equation}
  \label{eq:21}
  \int_0^{\tau}\d u(1 - e^{-i\omega u}) = \tau
  + \frac{1}{i\omega}\big(e^{-i\omega\tau} - 1\big)
\end{equation}
this implies that Eq.~(\ref{eq:12}) in the Fourier domain reads
\begin{equation}
  \label{eq:22}
  \widetilde{\nabla_{\perp}\psi}(\omega)
  = \frac{1 - \lambda_{\tau}(i\omega)}{\chi}\left(
    \chi\widetilde{\nabla_{\perp}\psi}(\omega)
    + \sqrt{2D_r^0}i\omega\tilde Z_{\omega}
  \right)
\end{equation}
which can easily be solved for
$\widetilde{\nabla_{\perp}\psi}(\omega)$ to yield Eq.~(\ref{eq:23}).

\section{Poles near the Origin}
\label{sec:poles-near-origin}

For $\chi < \chi_c$ we can rule out a pole at the origin. Then we may
rewrite the condition $\lambda_{\avr\tau}(s) = 0$ as
\begin{equation}
  \label{eq:46}
  s^2 - \chi\avr\tau s - \chi\big(e^{-s\avr\tau} - 1\big) = 0,
\end{equation}
and to third order in $s$
\begin{equation}
  \label{eq:47}
  \frac16\chi\avr\tau^3s^3 + (1 - \chi\avr\tau^2/2)s^2 = 0.
\end{equation}
This is solved by $s^* = -3(\chi_c/\chi - 1)/\avr\tau$.

\section{Details of the Numerics}
\label{sec:details-numerics}

We determined $\delta\varphi^2(t)$ using a numerical inverse Laplace
transformation of Eq.~(\ref{eq:28}) by the method of
\citet{abate+valko04} implemented in \texttt{Python} with the help of
the multi-precision library \texttt{mpmath}.\cite{mpmath} The
translational mean square displacement $\delta r^2(t)$, we determined
by numerical integration of Eq.~(\ref{eq:33}) using
\texttt{SciPy}'s\cite{scipy} \texttt{quad} method.

For the trajectories we used \texttt{SciPy}'s inverse fast Fourier
transform \texttt{ifft} to determine $\varphi(t)$ from
Eq.~(\ref{eq:24}) and then Euler integration of Eq.~(\ref{eq:8}).

%

\end{document}